\documentclass[prb,twocolumn,floatfix,showpacs]{revtex4}
\usepackage{graphicx}
\usepackage{times}
\newcommand{\be}{\begin{eqnarray}}
\newcommand{\ee}{\end{eqnarray}}

\newcommand{\ba}{\begin{array}}
\newcommand{\ea}{\end{array}}

\begin{document}
\title{Anderson transition in systems with chiral symmetry}
\author{Antonio M. Garc\'{\i}a-Garc\'{\i}a}
\affiliation{Physics Department, Princeton University, Princeton,
New Jersey 08544, USA}
\affiliation{The Abdus Salam International
Centre for Theoretical Physics, P. O. Box 586, 34100 Trieste, Italy}
\author{Emilio Cuevas}
\affiliation{Departamento de F\'{\i}sica, Universidad de Murcia,
E30071 Murcia, Spain}
\date{\today}

\begin{abstract}
Anderson localization is a universal quantum feature caused by
destructive interference. On the other hand chiral symmetry is a key
ingredient in different problems of theoretical physics: from
nonperturbative QCD to highly doped semiconductors. We investigate the
interplay of these two phenomena in the context of a three-dimensional
disordered system. We show that chiral symmetry induces
an Anderson transition (AT) in the region close to the band center.
Typical properties at the AT such as multifractality and critical
statistics are quantitatively affected by this additional symmetry.
The origin of the AT has been traced back to the power-law decay of
the eigenstates; this feature may also be relevant in systems
without chiral symmetry.
\end{abstract}
 \maketitle

The combination of global symmetries and dimensionality provides us
with a useful classification scheme for the study of Anderson
localization.\cite{anderson} According to the one-parameter scaling
theory (ST),\cite{one} all eigenstates of a disordered system are
exponentially localized in fewer than two dimensions in the
thermodynamic limit and undergo a localization-delocalization
transition [or Anderson transition (AT)] in higher dimensions, if
the strength of disorder is sufficiently weak. However, deviations
from ST predictions have been found in systems with additional
discrete symmetries. A typical example is a disordered lattice
without on-site disorder but random hopping amplitudes.\cite{dyson}
Formally, the system may be considered as two independent
sublattices with nonzero matrix elements connecting only sites of
different sublattices. The corresponding Hamiltonian is invariant
under a simultaneous sign flip of an arbitrary eigenvalue and the
eigenstate components of one of the associated sublattices. As a
consequence of this discrete symmetry, referred to as chiral
symmetry, eigenvalues of the Hamiltonian come in pairs of $\pm
\epsilon_i$, that is, the spectrum is invariant around the point
$\epsilon  = 0$. This symmetry is an important ingredient in
different problems of theoretical physics: from the QCD Dirac
operator\cite{SV} to  bosons in random media \cite{GC} and the
spectrum of a highly doped semiconductor.\cite{efros}

From the definition of chiral symmetry it is clear that one must
distinguish between eigenvalues near zero (the origin) and distant
from it (the bulk). In the bulk, spectral correlations in the
metallic limit are described by the universal results of random
matrix theory (RMT), usually referred to as Wigner-Dyson (WD)
statistics.\cite{WD} Weak-localization corrections to these universal
results are obtained by mapping the localization problem onto a
supersymmetric nonlinear $\sigma$ model.\cite{efetov} The
transition to localization occurring for stronger disorder is beyond
the reach of current analytical techniques. Numerical results
suggest that the AT is characterized by  multifractal
\cite{schreiber} eigenfunctions, a scale-invariant spectrum,
\cite{sko} and spectral correlations (usually referred to as
''critical statistics'' \cite{KMu}) different from Poisson and WD
statistics.\cite{sko,chi} These properties are supposed to be universal,
namely, they do not depend on the microscopic details of the
Hamiltonian but only on the dimensionality and symmetries of the
system.

In contrast, much less is known about the region close to the origin
where the chiral symmetry plays a crucial role. An exception is the
metallic limit, where the effect of this symmetry on level
statistics \cite{forrester} has been investigated in detailed with
the help of the powerful analytical techniques of RMT. Beyond this
region we are still far from even a qualitative understanding of the
interplay between disorder and chiral symmetry. As a general rule,
chiral symmetry tends to delocalize eigenstates close to the origin
since weak-localization corrections vanish.\cite{bro1} Indeed in
the one-dimensional case it can be rigorously proved
\cite{dyson} that, in disagreement with the ST prediction,
eigenstates at the band center are not localized exponentially. In
two dimensions, localization properties seem to be very sensitive to
the microscopic form of the random potential.\cite{dwv} For the
specific case of a weak Gaussian disorder, it is well established
that eigenvectors at the band center remain delocalized.\cite{gade}
In three-dimensions (3D), for weak disorder and time-reversal
invariance, the results of Refs.\onlinecite{evangelou,AJ,nikolic} suggest
that chiral symmetry may induce power-law localization of the zero-energy
eigenstates.

Surprisingly, with the exception of certain one-dimensional systems
with-long range disorder,\cite{AK,ant1} the transition to
localization in systems with chiral symmetry has hardly been studied
despite its potential applications to the description of the chiral
phase transition in QCD,\cite{ant12} or the metal-insulator
transition in highly doped semiconductors. The present Brief Report is a
step in this direction. We investigate the interplay between chiral
symmetry and localization in a three-dimensional system with
short-range disorder. Our main finding is that chiral symmetry
induces an AT in the region close to the origin. Eigenstates are
power-law localized with an exponent that increases as we move from
the origin.  The AT occurs when this exponent matches the
dimensionality of the space. We recall that, since the AT is to a
great extent universal, the validity of our results does not depend
on the microscopic details of the model but only on the
dimensionality of the space and the (chiral) symmetry of the
Hamiltonian.

We consider the following 3D tight-binding Hamiltonian with only
off-diagonal disorder:
\begin{equation}
\label{ourmodel} {\cal H}=\sum_{\langle ij \rangle
}[t_{ij}{\text{e}}^{\text{i}\theta_{ij}}
  a^{\dag }_i a_j+\text{H.c.}]\;,
\end{equation}
where the sum is restricted to nearest neighbors and the operator
$a_i \; (a^{\dag }_i)$ destroys (creates) an electron at the $i$th site
of the 3D cubic lattice. We break time-reversal invariance by
introducing a random  magnetic flux described by the above Peierls
phase $\text{e}^{i\theta_{ij}}$ with $\theta_{ij}$ uniformly
distributed in the interval $[-\pi,\pi]$. Although we mainly focus
on the broken time-reversal case, we have also checked that our main
conclusions are valid if time-reversal invariance is preserved. The
hopping integrals $t_{ij}$ are real random variables satisfying the
probability distribution
\begin{eqnarray}
{\cal P}(\ln t_{ij}) = 1/W \, \, \, \, \, \text{  for  } -W/2 \leq
\ln t_{ij} \leq W/2\;,
\end{eqnarray}
and zero otherwise. The above exponential distribution provides an
effective way of setting the strength of off-diagonal disorder.\cite{off}

In order to proceed, we compute eigenvalues and eigenvectors of the
Hamiltonian (\ref{ourmodel}) for different volumes $L^3$ by
using standard numerical diagonalization techniques. The number of
disorder realizations for each volume $L^3$ ranges from $10^4$
($L=10$) to $500$ ($L=20$). Hard-wall boundary conditions are
imposed in order to better examine the asymptotic decay of
eigenstates in real space (see below). Since we are interested in
the effect of chiral symmetry we mainly focus on a small energy
window close to the origin $\epsilon = 0$. The eigenvalues thus
obtained are appropriately unfolded, i.e., they were rescaled so that
the spectral density on a spectral window comprising several level
spacings is unity. 

Our first task is to look for a mobility edge in a small spectral
window close to the origin. The chosen interval, $\epsilon \in
(10^{-4},10^{-3})$ in our case, is not important provided that it is
close to the origin. For other intervals one gets similar results
but for a different critical disorder. We recall that in general the
value $W = W_c$ for which the system undergoes an AT is quite
sensitive to the details of the Hamiltonian and consequently is not
universal.  In order to locate the mobility edge we use the finite
size scaling method.\cite{sko} First we evaluate a certain spectral
correlator for different sizes and disorder strengths $W$. Then we
locate the mobility edge by finding the disorder $W_c$ such that the
spectral correlator analyzed becomes size independent.\cite{sko}

In our case we investigate the level spacing distribution $P(s)$
[the probability of finding two neighboring eigenvalues at a distance
$s_i = (\epsilon_{i+1} - \epsilon_{i})/\Delta $, with $\Delta$ being
the local mean level spacing]. The scaling behavior of $P(s)$ is
examined through the following function of its variance:\cite{C99}
\begin{equation}
\eta (L,W) = [{\rm var} (s)- {\rm var_{WD}}\,] / [\,{\rm
var_{P}}-{\rm var_{WD}}] \;, \label{eta}
\end{equation}
which describes the relative deviation of ${\rm var} (s)$ from the
WD limit. In Eq. (\ref{eta}) ${\rm var} (s)=\langle
s^2\rangle-\langle s \rangle^2$, where $\langle \cdots \rangle$
denotes spectral and ensemble averaging, and ${\rm var_{WD}} =
0.286$ and ${\rm var_P} = 1$ are the variances of WD and Poisson
statistics, respectively. Hence $\eta = 1 (0)$ for an insulator
(metal). Any other intermediate value of $\eta$ in the thermodynamic
limit is an indication of a mobility edge.

\begin{figure}
\includegraphics[width=1.0\columnwidth,clip]{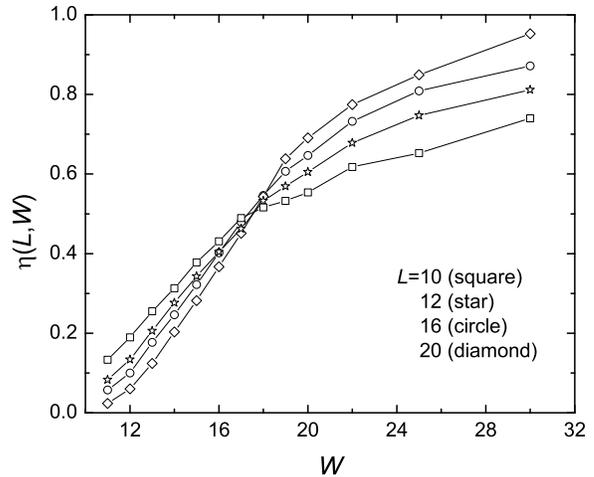}
\caption{Scaling variable $\eta$ as a function of disorder $W$ for
different volumes; only eigenvalues in the window $10^{-4} \le
\epsilon \le 10^{-3}$ have been considered. The system undergoes an
AT at $W = W_c \sim 18 \pm 1$. In all cases the statistical error
(not shown) is $\sim 0.003$.} \label{fig29}
\end{figure}

In Fig. 1 we plot the $W$ dependence of $\eta$ for different system
sizes. It is clear that, in the window of energy studied $\epsilon
\in (10^{-4},10^{-3})$, the mobility edge signaling an AT is located
around $W_c \sim 18$. We have also found that $W_c$ increases as the
energy window of interest gets closer to the origin. For a weaker
(stronger) disorder, $\eta$ tends slowly to the WD (Poisson) results.
This slow convergence to WD statistics (see Fig. 1) suggests that
eigenstates may still have some kind of structure even on the
delocalized side of the transition. The analysis of eigenfunctions
will show that this is the case. We have checked that the use of scaling
variables other than $\eta$ \cite{PV92,sko} do not alter the
results. For the sake of completeness we repeat the calculation for
the case of time-reversal invariance, $\theta_{ij} = 0$ in Eq.
(\ref{ourmodel}). We have also observed a mobility edge with similar
properties but at a slightly weaker disorder $W_c \sim 15$, in
agreement with previous results for the standard 3D AT.\cite{hof}
Finally, we remark that the study of $\eta(W,L)$ for different 
disorders and system sizes is equivalent to the standard renormalization
group analysis of the dimensionless conductance $g$. For instance, according
to the one-parameter scaling theory, $g$ should be scale invariant at the AT.
Any spectral correlator, $\eta$ in particular, is a function of $g$ only and,
consequently, it should be scale invariant at the AT as well. We have chosen
$\eta$ instead of $g$ for numerical reasons. The former is easier to calculate
and provides much more stable results.      

We now study  whether level statistics around the
mobility edge are compatible with those of an AT. Spectral
correlations at the AT (critical statistics) combine typical
properties of a metal (WD) with those of an insulator (Poisson).
Thus level repulsion $P(s) \rightarrow 0$ for $s \rightarrow 0$,
typical of WD statistics is still present at the AT. However, a
long-range correlator such as the the number variance
$\Sigma^2(\ell)=\langle (N_\ell -\langle N_\ell \rangle)^2
\rangle \sim \chi \ell$ for $\ell \gg 1$ ($N_\ell$ is the number of
eigenvalues in an interval of length $\ell$) is asymptotically
linear,\cite{chi} as for an insulator $\Sigma^2(\ell) \sim \ell$, but
with a slope $\chi < 1\,$ ($\sim 0.27$ for 3D AT).

\begin{figure}
\includegraphics[width=1.0\columnwidth,clip]{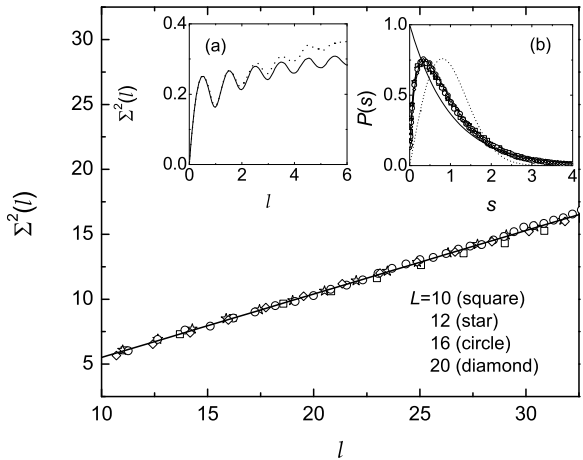}
\caption{The $l \gg 1$ behavior of the number variance $\Sigma^2(l)$
and $P(s)$ [inset (b)] in the spectral window $\epsilon \in
(10^{-4},10^{-3})$ at the critical point $W_c \sim 18$ for different
volumes. $\Sigma^2(l)$ is linear with a slope (solid line) $\chi
\sim 0.49 \pm 0.01$, $P(s) \sim s$ for $s \ll 1$, and both are
size independent. These features are typical signatures of critical
statistics. For $W = 0.2 \ll W_c$ [inset (a)] the number variance
(dotted line) is well described by the prediction  of chiral RMT
(solid line).} \label{fig31}
\end{figure}

As shown in Fig. 2 all these features are also present in our model.
However, we have found that the slope of the number variance $\chi
=  0.49 \pm 0.01$ is roughly twice that of the standard 3D AT. We refer to
the analysis of eigenvectors (see below) for a qualitative
explanation of this feature.

We have verified that the number variance in the limit of weak
disorder agrees with the prediction\cite{forrester} of chiral RMT 
[see inset (a) Fig. 2]. For weak enough disorder results depend on
whether the time-reversal invariance is broken [inset (a) in Fig. 2]
or not (not shown).  As a general rule, as disorder is increased the
magnetic flux gets weaker, for $W \geq 10$ we could not observe any
difference. Additionally, we have found that, if chiral symmetry is
broken, by for instance adding a weak diagonal disorder, level
statistics are very close to the Poisson limit typical of an insulator.

To conclude, the analysis of level statistics shows that, as a
consequence of the chiral symmetry of the model, there is an AT
close to the band center with properties similar
but not equal to those of the standard 3D AT.

One of the signatures of an AT transition is the multifractality of
the eigenstates. An eigenfunction is said to be multifractal (for an
alternative definition see Ref. \onlinecite{cuevas}) if the eigenfunction
moments $P_q=\int d^d r |\psi(r)|^{2q}\propto L^{-D_q(q-1)}$ present
anomalous scaling with respect to the sample size $L$, where $D_q$
is a set of different exponents describing the transition.\cite{wegner,mirlin}

After performing ensemble and spectral averaging in the spectral
window $\epsilon \in (10^{-4},10^{-3}$) we have found that the
eigenfunctions are indeed multifractal ($D_2 = 0.69 \pm 0.02$, $D_3 =
0.45 \pm 0.01$, and $D_4 = 0.340 \pm 0.007$). These values of $D_q$ are,
roughly speaking, a factor of $2$ smaller than the ones at the standard 3D
AT. This is consistent with our results from level statistics, where a
similar factor was found for the slope $\chi$ of the number variance
$\Sigma^2(\ell) \sim \chi \ell$. Qualitatively, this numerical
difference can be traced back to the chiral symmetry of our model:
according to the ST, parameters such as $\chi$ or $D_q$ are
functions of the dimensionless conductance $g_c$ at the AT. Although
an explicit expression is not known, it is expected that, at least
for sufficiently high dimensions, both $D_2$ and $1 -\chi$ will be
proportional to $g_c$. On the other hand, it is well established
\cite{gade,bro1} that weak-localization corrections vanish in
systems with chiral symmetry. Hence, a stronger disorder is needed
to reach the AT region. As a consequence, the chiral $g_c$ will be
smaller, in qualitative agreement with our results. Having discussed
the details of the AT close to the origin, we now clarify the
physical mechanism causing this transition. Chiral systems do not
have weak-localization corrections, so the transition to
localization must be in part due to some nonperturbative effect,
such as tunneling combined with the progressive weakening of the
effect of the chiral symmetry as we move away from the origin. On
the other hand, the slow rate of convergence toward WD statistics
observed in Fig. 1 suggests that, although eigenstates seem to be
delocalized in the thermodynamic limit, they still have some kind
of localization center. This is in agreement with the results of
Ref. \onlinecite{evangelou}, where it was found that the zero-energy
wave function of a time-reversed, weakly disordered 3D chiral system
was power-law localized $\psi(r) \sim  1/r^{\alpha(W)}$. We recall
that, according to Refs. \onlinecite{mirlin} and \onlinecite{levitov},
power-law localization induces an AT in any dimension provided that
the decay exponent $\alpha$ matches the dimensionality $d$ of the space.
If $\alpha < (>)\; d$, eigenstates tend to be delocalized (localized)
in the thermodynamic limit but the degree of convergence is
slow.\cite{cuevasor}

Based on the above facts we claim that the transition to
localization close to the origin observed in the Hamiltonian
(\ref{ourmodel}) has its origin in the power-law localization of the
eigenstates. We remark that this mechanism should be at work for any
system with chiral symmetry and disorder such that a mobility edge
appears close to the origin. The Hamiltonian (\ref{ourmodel})
was chosen for the sake of simplicity. It is just one of the
simplest representatives of the universality class associated with
the AT in systems with chiral symmetry.

Moreover, for a given spectral window close to the origin, we expect
the exponent $\alpha$ describing the decay to increase with the
disorder strength $W$. The AT will occur for a $W = W_c$ so that
$\alpha \sim d = 3$. For stronger disorder $W > W_c$ eigenstates
must be eventually exponentially localized, though they may still
possess a power-law tail for distances smaller than the localization
length.

We have tested this conjecture by analyzing the typical decay of
eigenstates in the window $10^{-4} \le \epsilon \le 10^{-3}$ for $L
= 20$ and different $W$'s (similar results are obtained if $W$ is
kept fixed and the energy window is modified) through the following
correlation function:\cite{IT94}
\begin{equation}
g(r)=\langle \ln { (|\psi_{j_{\rm max}}|/|\psi_j|)} \rangle\, , \; \;
r=|{\bf r}_j-{\bf r}_{j_{\rm max}}|\, ,
\end{equation}
where $j_{\rm max}$ denotes the lattice site with the largest amplitude
and the angular brackets stand for the ensemble average. In order to
examine the asymptotic decay of $g(r)$ (we restrict ourselves to
the range $8 \le r \le 22$), we fit the numerical results to a curve
of the form $g(r)=\ln (A r^{\alpha} + B)$ with $A$, $B$, and $\alpha$
fitting parameters. The parameters $A$ and $B$ describe finite-size
effects and remnants of the eigenstate core (nondecaying) part. The
exponent $\alpha$ (depicted in the inset of Fig. 3) provides us with
valuable information about the decay of the eigenstates.

\begin{figure}
\includegraphics[width=1.0\columnwidth,clip]{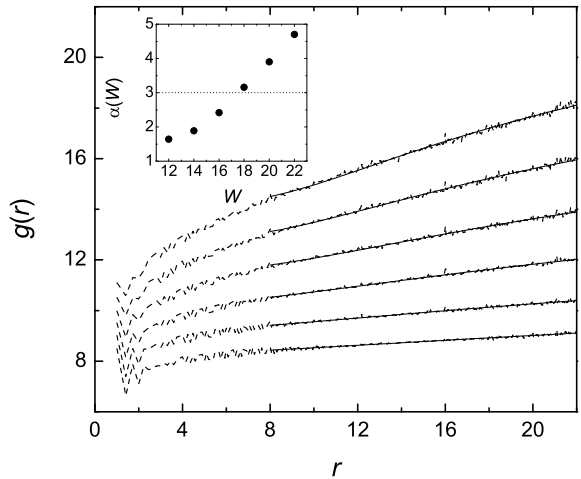}
\caption{The correlation function $g(r)$ as a function of $r$
(dashed lines) for a system size $L=20$ and $W=$ 12, 14, 16, 18, 20,
and 22 from bottom to top. The numerical curves were fitted to a
curve of the form $g(r)=\ln (A r^{\alpha} + B)$ (solid lines). The
best-fitting parameter $\alpha$ is represented as a function of $W$
in the inset. In all cases the statistical error (not shown) is
$\sim \alpha/50$.} \label{fig32}
\end{figure}

The excellent agreement (see Fig. 3) between the numerical results
and the fitting curve indicates that eigenstates are power-law
localized. Moreover, the exponent $\alpha$ controlling the decay
increases with $W$ and tends to the conjectured value $\alpha = 3$
as we get close to the critical disorder $W_c \sim 18$. Thus chiral
symmetry combined with strong disorder induces power-law
localization and eventually an AT close to the origin. This is the
most relevant result of the paper.

We remark that the relation between chiral symmetry and power-law
localization has already been established in the context of QCD (see
Ref. \onlinecite{AJ} and references therein). Specifically, it was
found that the low-lying eigenstates of the QCD operator (with gauge
configurations given by the instanton liquid model) are power-law
localized due to the long-range behavior of certain nonperturbative
solutions (instantons) of the classical equations of motion. We
speculate that a similar mechanism is at work in our case.

A final comment is in order: the claim that the power-law
localization is the precursor of the AT is not in contradiction
with the multifractal structure of the eigenstates. Since $g(r)$
is averaged over many realization of disorder, all multifractal
fluctuations are washed out.  What remains is the smooth skeleton
of the eigenstate which, in our case, has a power-law tail.


To conclude, we have investigated the interplay between chiral
symmetry and Anderson localization in a 3D Anderson model with
off-diagonal disorder. Close to the origin we have found a mobility
edge induced by the chiral symmetry with properties quantitatively
different from those of a standard 3D AT. This chiral AT can be traced
back to the power-law nature of eigenstates close to the origin. Our
results are relevant to chiral systems undergoing a metal-insulator
transition as a function of disorder. Typical examples include the
QCD Dirac operator around the chiral phase transition and highly
doped semiconductors where off-diagonal bond
disorder plays a dominant role.\\

A.M.G. acknowledges support from the European Union,
Marie Curie program, Contract No. MOIF-CT-2005-007300.
E.C. thanks the FEDER and the Spanish DGI for financial
support through Project No. FIS2004-03117.


\end{document}